\documentclass[%
 preprint,
 amsmath,amssymb,
 aps, physrev,
]{revtex4-2}
\usepackage{graphicx}
\usepackage{dcolumn}
\usepackage{bm}
\usepackage[utf8]{inputenc}
\usepackage{amsmath}
\usepackage{amsthm}
\usepackage{amssymb}
\usepackage{bbm}
\usepackage{braket}
\usepackage{tikz-cd}
\usepackage{tensor}
\usepackage{color}
\usepackage{dsfont}
\usepackage{hyperref}
\usepackage{tabularx}

\newcommand{\of}[1]{\left( {#1}\right)}

\DeclareMathOperator{\im}{Im}

\begin{document}

\preprint{APS/123-QED}

\title{\textbf{Non-Abelian fractional Chern insulators from an exactly solvable two-body model} 
}%

\author{Joseph R. Cruise}
    \email{Contact author: c.joseph@wustl.edu}
\author{Alexander Seidel}
\affiliation{%
  Department of Physics, Washington University in St. Louis, St. Louis Missouri 63130, USA 
}%
\date{\today}

\begin{abstract}
\end{abstract}
\begin{abstract}
We construct a class of lattice Hamiltonians whose single-particle spectrum consists of an arbitrary number of exactly degenerate flat bands that reproduce the analytic structure of the first $p$ Landau levels restricted to the lattice. When combined with local bosonic contact interactions, these models become exactly solvable frustration-free parent Hamiltonians for FCIs that realize both Abelian and non-Abelian parton quantum Hall states.
Using exact diagonalization, we confirm the expected zero-mode counting for variants of the model stabilizing the bosonic Jain-$21$ state as well as the non-Abelian $22-$ and $33$-states, which are expected to support Ising- and Fibonacci-type anyons, respectively.
Our construction provides an exactly solvable lattice realization of multi–Landau-level physics and offers a new framework for studying FCIs with Chern number $\mathcal{C}>1$. More broadly, it supplies a family of idealized lattice models that capture the analytic structure of continuum Landau levels while remaining compatible with exponentially local hopping.
\end{abstract}
\maketitle


\section{\label{sec:intro}Introduction}
Chern insulators (CIs) are generalizations of quantum Hall (QH) states that owe their quantized hall conductivity to the topology of their band structure. As Haldane demonstrated \cite{haldane_model_1988}, a CI does not require a net external magnetic field, as long as some other mechanism breaks time reversal symmetry in a topologically non-trivial way. 
Indeed, research on the geometric origins of Chern insulators (CIs), particularly fractionally filled CIs (FCIs) \cite{parameswaran_fractional_2013,bergholtz_topological_2013,neupert_fractional_2011,tang_high-temperature_2011,sun_nearly_2011}, suggests that model FQH systems can be understood as special points in a broader CI phase space.
Here, 
a strong external magnetic field produces uniform momentum-space Berry curvature and introduces a magnetic length scale that replaces the lattice spacing, thereby restoring Galilean symmetry. 
Such an identification underlies several schemes for interpolation between FQH and FCI models \cite{lee_band_2016,roy_band_2014,lee_band_2017,mera_engineering_2021,dong_exact_2020,qi_generic_2011,claassen_position-momentum_2015}, and rigorous mathematical results exploring the relationship are currently the subject of intense research \cite{wang_exact_2021,estienne_ideal_2023,sarkar_ideal_2025,mera_kahler_2021,wang_origin_2023,sarkar_symmetry-based_2025,liu_theory_2025,mera_uniqueness_2024}.

Although the relationship between FQH systems and FCIs continues to be refined at a formal level, on a practical level the search for realizable models has produced a coherent design philosophy. In the absence of a magnetic length scale one must reckon with an underlying lattice, which obscures the familiar picture of FQH trial states as unique incompressible ground states of local pseudopotential parent Hamiltonians \cite{trugman_exact_1985,haldane_pseudopotential}. In particular, it has been rigorously established that no strictly local lattice model can host perfectly flat topological bands \cite{chen_impossibility_2014,read_compactly_2017}. This result sharply delineates the constraints faced by lattice band models for fractionalized topological phases, implying that viable constructions must either retain band dispersion or allow for exponentially decaying hopping amplitudes.
 Indeed, both dispersive-band models \cite{trescher_flat_2012,wang_nearly_2011,sun_nearly_2011,ma_spin-orbit-induced_2020,yang_topological_2012} and models with exponentially decaying hoppings \cite{kapit_exact_2010,glasser_exact_2015,liu_nonabelian_2013,kapit_three-_2013} are known to support FCIs, often in the same topological phases as counterpart continuum FQH models.

In the following, we will study lattice models that support fractionalized phases with Abelian as well as non-Abelian topological order. Historically, a distinct route to exact solvability for such phases has been provided by lattice constructions based on commuting-projector or gauge-theoretic mechanisms. Canonical examples include Kitaev-type spin models and Levin--Wen string-net models \cite{kitaev_anyons_2006,levin_string_2005} which realize a wide range of (potentially non-Abelian) topological quantum field theories directly on the lattice. In these settings, exact solvability is a consequence of an extensive set of conserved quantities or local projector constraints, often involving multi-spin interactions, enlarged local Hilbert spaces, or effective descriptions that go beyond simple two-body Hamiltonians. While these models have been central to establishing the conceptual framework of topological order and the bulk--boundary correspondence, their route to exact solvability differs fundamentally from the band-based or zero-mode constructions considered here.

The present work pursues a complementary and substantially more constrained route to exact solvability. Rather than relying on commuting-projector structures or enlarged local Hilbert spaces, we seek lattice Hamiltonians built solely from local two-body interactions whose low-energy structure is determined by zero-mode constraints. In this setting, exact solvability does not arise from an extensive set of conserved projectors, but instead emerges from the analytic properties and mutual organization of the underlying single-particle states. While related zero-mode constructions have long been central to exactly-solvable models in the continuum FQH setting, their extension to lattice systems has proven to be considerably more subtle. On the one hand, the connection between lattice FCI models and continuum FQH parent Hamiltonians has been actively explored and exploited in a variety of contexts \cite{regnault_fractional_2011}, and recent work has clarified and systematized the role of pseudopotentials beyond the lowest Landau level \cite{ledwith_vortexability_2023}. On the other hand, systematic lattice constructions that preserve entire zero-mode spaces while remaining local—particularly in the presence of non-Abelian topological order—remain incomplete. The approach developed in this work enables the construction of exactly solvable lattice models with ultra-short ranged two-body interactions and non-Abelian zero-mode structure, which may serve as concrete targets for engineered (bosonic) Hamiltonians.

The fact that continuum FQH parent Hamiltonians capitalize on the analytic clustering properties of special many-body wavefunctions makes a direct transposition of such models to the lattice highly nontrivial, particularly if the goal is to preserve the entire zero-mode space. For the Abelian $\nu=1/2$ Laughlin state, however, a celebrated work by Kapit and Mueller (KM) \cite{kapit_exact_2010} demonstrates that precisely this can be achieved. The KM model is an exponentially local extension of the Harper--Hofstadter model \cite{hofstadter}, featuring an extensively degenerate ground-state manifold whose analytic structure reproduces that of the lowest Landau level (LLL), restricted to a lattice. In this setting, a repulsive contact interaction plays a role analogous to the $V_0$ Haldane pseudopotential in the continuum, rendering the $\nu=1/2$ Laughlin state an exact zero-energy eigenstate.

A central limitation of bosonic lattice constructions based on  two-body contact interactions is apparent when one is restricted to the LLL: in this case, such interactions uniquely stabilize the $\nu=1/2$ Laughlin state. Generalizations of this principle within the LLL are nevertheless well understood in the continuum, where a broad class of exactly solvable parent Hamiltonians can be formulated within a ``zero-mode paradigm.'' In this framework, local interactions enforce analytic clustering constraints on many-body wavefunctions projected to the LLL, giving rise to characteristic zero-mode spaces. In rotationally invariant geometries, the counting of these zero modes matches that of an edge conformal field theory, thereby encoding the associated topological quantum field theory through bulk--boundary correspondence \cite{MOORE1991362, read_conformal_2009}. Known realizations of this paradigm include trial states defined through conformal blocks and, in many cases, equivalently through Jack polynomials \cite{macdonald_symmetric_1995,bernevig_model_2008,bernevig_generalized_2008}. The corresponding zero-mode spaces are further organized by generalized Pauli principles, which encode their counting in terms of constrained occupation-number patterns \cite{haldane_fractional_1991,bernevig_generalized_2008,bandyopadhyay_entangled_2018}.

Along these lines, sophisticated extensions of Laughlin-state parent Hamiltonians are well known in the continuum. Historically, such developments have focused primarily on LLL-projected states, where construction principles exploit the fact that analytic clustering properties of many-body wavefunctions translate into constraints on symmetric polynomials that are tractable within commutative algebra, and where field-theoretic interpretations are naturally available through the Moore--Read framework \cite{MOORE1991362}. However, these extensions typically require parent Hamiltonians that go beyond simple two-body contact interactions.

On the other hand, it has been known for some time that the same ultra-local two-body interactions which stabilize Laughlin states in the LLL may stabilize more general ``parton states'' when projected onto the first $p$ Landau levels, with $p>1$ \cite{jain_incompressible_1989, wen_non-abelian_1991,wu_non-abelian_2017,balram_parton_2018,balram_fractional_2018,balram_parton_2019,kim_even_2019,balram_fractional_2020,fuagno_non-abelian_2020,balram_abelian_2021,balram_non-Abelian_2021,balram_parton_2021,balram_nature_2022,faugno_unconventional_2021,timmel_nonabelian_2023}. The parton construction introduces effective flavor degrees of freedom which, upon gauging, generically give rise to non-Abelian topological orders \cite{wen_projective_1999,zhang_establishing_2013,anand_realspace_2022}. At the same time, a rigorous understanding of the full zero-mode structure of such multi–Landau-level continuum models cannot proceed by a straightforward extension of LLL methods, and has become possible only quite recently through the development of second-quantized approaches \cite{bandyopadhyay_entangled_2018,bandyopadhyay_local_2020, tanhayi_ahari_partons_2023, cruise_sequencing_2023}.

With the structure of multi–Landau-level parton parent Hamiltonians in the continuum now under quantitative control, the remaining obstacle to extending constructions such as the KM model lies at the level of single-particle physics. In particular, one must engineer lattice models in which the first $p$ Landau levels are realized as {\em degenerate} flat bands (with $p=1$ for the original KM model). Such a setting is naturally suited to parton FCIs and their parent Hamiltonians, and closely resembles the continuum limit of Landau levels in rhombohedral graphene, which has lately attracted significant interest \cite{wu_non-abelian_2017,henck_flat_2018,park_topological_2023,seifert_increasing_2024,zhou_fractional_2024,zhang_corelated_2024}. Furthermore, recent work suggests that the novel setting of degenerate Landau levels may generically allow for improved many-body gaps through the mechanism of pseudopotential engineering \cite{zhang_2025_lowest}. 

In this work, we demonstrate a method to generalize the KM construction to produce an arbitrary number of degenerate Landau levels on a lattice, while retaining exponentially decaying single-particle hoppings. This construction yields exactly solvable parent Hamiltonians for (bosonic) non-Abelian FCI states.
 We emphasize that while the connection between non-Abelian FCI physics and bands with Chern number $\mathcal{C}>1$ is well established \cite{zhang_establishing_2013,wu_bloch_2013,sterdyniak_series_2013,wu_fractional_2015,behrmann_model_2016,wang_origin_2023}, to the best of our knowledge this is the first realization of such topological orders arising from ultra-short ranged two-body interactions within an exactly solvable lattice framework.

Beyond their intrinsic theoretical appeal, FCIs are also of considerable interest from a broader physical perspective, as they may be generically more robust than their FQH counterparts. In the absence of a strong, quantizing magnetic field, the relevant gap scale $\Delta$ is set by the inverse lattice constant $a^{-1}$, which typically leads to energy scales $\Delta_{\mathrm{FCI}} \gtrsim \Delta_{\mathrm{FQH}}$. As a result, FCIs not only obviate the extreme magnetic fields required to observe the FQH effect, but may also persist at substantially higher temperatures. 
Indeed, many of these expectations have been borne out by a striking degree of success in both theory and experiment, particularly in moiré materials \cite{abouelkomsan_particle-hole_2020,park_topological_2023,crepel_chiral_2024,makov_flat_2024,wang_higher_2025} as well as in quantum-simulator platforms. In this context, we hope that the models introduced here may serve as a useful starting point for the analysis of FCIs and flat-band physics in twisted transition-metal dichalcogenides \cite{redekop_direct_2024,kang_evidence_2024}, rhombohedral stacks of graphene \cite{zhang_corelated_2024,lu_fractional_2024,zhang_moire_2025,xie_tunable_2025,waters_chern_2025,han_signatures_2025}, and as concrete targets for realizing non-Abelian physics in cold-atom or photonic quantum simulators \cite{tai_microscopy_2017,palm_growing_2024,yao_realizing_2013,cooper_reaching_2013,leonard_realization_2023,clark_observation_2020,wang_realization_2024}.

The rest of the paper is organized as follows. In Sec.~\ref{latticeLL}, we will review the details of Landau levels on a lattice with periodic boundary conditions, and then introduce our model in Sec.~\ref{onebod}. In Sec.~\ref{results} we will present numerical results in support of a non-Abelian ground state degeneracy, and finally we will conclude in Sec.~\ref{conc}.

\section{Landau levels on a lattice \label{latticeLL}}

In this section, we lay some groundwork for a lattice Hamiltonian with degenerate flat bands that are identical to the first $p$ Landau levels on a torus, restricted to the lattice sites.
To this end, we construct continuum Landau-level wavefunctions whose
translation properties are compatible with the lattice, so that upon
restriction to sites they define magnetic Bloch states on the lattice.

Consider a square lattice of integer dimensions $M\times N$ with twisted periodic boundary conditions and subject to a constant magnetic field with $\phi$ flux quanta per plaquette. Consistency of the magnetic periodic boundary conditions requires that the total flux $MN\phi$ equals an integer $L$.
Adjacent sites on the lattice are connected by magnetic translation operators $T_x$ and $T_y$ which for $R = (x,y)$ satisfy
\begin{align}
    T_x c^\dagger_R T_x^\dagger &= e^{i\theta(R,R+\hat x)}c^\dagger_{R+\hat x} \nonumber\\ 
    T_y c^\dagger_R T_y^\dagger &= e^{i\theta(R,R+\hat y)}c^\dagger_{R+\hat y},
\end{align}
and boundary conditions are handled via the identification
\begin{align}\label{MPBC}
    c^\dagger_R = e^{-i\phi_1}e^{i\sum_{k=0}^{M-1}\theta(R+k\hat x, R+(k+1)\hat x)} c^\dagger_{R+M\hat x} \nonumber\\
    c^\dagger_R = e^{-i\phi_2}e^{i\sum_{k=0}^{N-1}\theta(R+k\hat y, R+(k+1)\hat y)} c^\dagger_{R+N\hat y}.
\end{align}
for $\phi_1$ and $\phi_2$ arbitrary solenoid fluxes \cite{niu_quantized_1985}. Here 
\begin{equation}
    \theta(i,j) = \int_j^i \vec A(\vec r) \cdot d\vec r 
\end{equation}
is the Peierls phase, with $A$ the magnetic vector potential. Note that the loop $C = T_x^\dagger T_y^\dagger T_x T_y$ satisfies 
\begin{equation}
    C c_i^\dagger C^\dagger = \exp\of{ i \oint \vec A \cdot d\vec r } c_i^\dagger = e^{-i2\pi \phi} c_i^\dagger
\end{equation}
which is gauge invariant.
In particular, $\phi=L/(MN)={\sf p}/{\sf q}$ is rational, with $\sf p$ and $\sf q$ co-prime.
We may choose integers $m, n, \mu$ and $\nu$ such that
$M = \mu m$, $N = \nu n$, and ${\sf q}=mn$. Then a magnetic unit-cell of dimensions $m\times n$ has $\sf p$ flux quanta, and $T_x^m$ and $T_y^n$ commute.

In the following sections, we will construct a class of 
 magnetic translation invariant Hamiltonians such that
\begin{equation}
     T_xHT_x^\dagger = H = T_y H T_y^\dagger.
\end{equation}
Eigenstates 
will be given in the form of Bloch-type wavefunctions satisfying 
\begin{align}
    H \psi_d(x,y) &= \epsilon_d(\vec k) \psi_d(x,y)\\
    T_x^m \psi_d(x,y) &= e^{-i k_x} \psi_d(x,y) \\
    T_y^n \psi_d(x,y) &= e^{-i k_y} \psi_d(x,y),
\end{align}
with $d$ a band index. For the moment we will restrict our attention to the case of a single band and drop the index $d$. In the language of first quantization, the magnetic translation operators act via
\begin{align*}
    T_x \psi(x,y) &= e^{i\theta(R-\hat x,R)} \psi(x-1,y)\\
    T_y \psi(x,y) &= e^{i\theta(R-\hat y,R)} \psi(x,y-1),
\end{align*}
on wavefunctions, resulting in the 
following conditions for magnetic Bloch states:
\begin{align}
    \psi(x,y) = e^{ik_x} e^{i\sum_{k=0}^{m-1}\theta(R-(k+1) \hat x,R-k\hat x)} \psi(x-m,y) \\
    \psi(x,y) = e^{ik_y} e^{i\sum_{k=0}^{n-1}\theta(R-(k+1) \hat y,R-k\hat y)} \psi(x,y-n). 
\end{align}
These equations are easiest to solve in Landau gauge, $A = \phi x \hat y$, where we find 
\begin{align}
    \psi(x,y) &= e^{ik_x}  \psi(x-m,y) \\
    \psi(x,y) &= e^{ik_y} e^{-i 2\pi \phi  n x} \psi(x,y-n).
\end{align}
Since Landau levels on the torus are invariant under the magnetic translations defined above, we may choose Landau level bases that conform to these conditions.
Specializing, for now, to the LLL, solutions can be given in terms of the elliptic theta functions with characteristic \cite{mumford_tata_1982,haldane_periodic_1985, read_quasiholes_1996}:
\begin{equation}
    \vartheta_{a,b}(z|\tau) = \sum_{l=-\infty}^\infty e^{i\pi(l+a)^2\tau + i 2\pi (l+a)(z+b)},
\end{equation}
for $\im \tau >0$, and $a,b \in \mathbb{R}$.  Letting $\tau = in/m$, we define the un-normalized wavefunction 
\begin{equation}\label{llstates}
    \Psi_{a,b}(x,y) = \vartheta_{a,b}\of{\left.\frac{z}{m}\right|\tau}e^{-\pi \phi y^2}.
\end{equation}
One may check that this function satisfies 
\begin{align}
    T_x^m \Psi_{a,b}(x,y) &= e^{-i2\pi a} \Psi_{a,b}(x,y) \label{tx_eig_vals}\\
    T_y^n \Psi_{a,b}(x,y) &= e^{i2\pi b} \Psi_{a,b}(x,y),
    \label{ty_eig_vals}
\end{align}
so that the values $a$ and $b$ label the momentum eigenvalues, the exact value of which can be fixed by imposing the full torus boundary conditions 
\begin{align}
    T_x^M \Psi_{a,b}(x,y) &= e^{i\phi_1} \Psi_{a,b}(x,y) \\
    T_y^N \Psi_{a,b}(x,y) &= e^{i\phi_2} \Psi_{a,b}(x,y),
\end{align}
for $\phi_1$ and $\phi_2$ the solenoid fluxes. It follows that 
\begin{align}
    a &= (- \alpha - \phi_1)/\mu \\
    b &= (\beta + \phi_2)/\nu
\end{align}
for integers $\alpha$ and $\beta$ with $ 0 \leq \alpha, \beta < \mu,\nu$.

It is worth remarking that our Landau-level functions differ from the standard
torus conventions used in much of the fractional quantum Hall literature
\cite{haldane_periodic_1985,read_quasiholes_1996}, where the half-period ratio
$\tau$ is fixed by the physical dimensions of the torus,
$\tau = iN/M$.
Landau levels on a torus must take the form of a
quasi-periodic 
function times a prescribed Gaussian factor such that the entire wavefunction has the periodicity of the underlying manifold: any choice of fundamental domain for the elliptic part of the wavefunction that satisfies this condition is equally valid. Identifying the fundamental domain with the magnetic unit-cell greatly simplifies the spectral analysis of the magnetic translation operators via (\ref{tx_eig_vals},\ref{ty_eig_vals}).

Since the operators $T_x^m$ and $T_y^n$ act unitarily on the space of lattice wavefunctions, (\ref{llstates}) gives an orthogonal basis for the (continuum) LLL. To obtain similar states in higher Landau levels, note that the operator
\begin{equation}
    \hat a = -\partial_{\bar z} - i \pi \phi y
\end{equation}
satisfies $\hat a\Psi_{a,b} = 0$ for all $a$ and $b$, and consider its Hermitian conjugate, the Landau level raising operator ${\hat a}^\dagger = \partial_z + i \pi \phi y$. One may show that ${\hat a}^\dagger$ preserves the eigenvalues of the magnetic translation operators. In practice, it is difficult to compute powers of ${\hat a}^\dagger$, since $\partial_z$ and $y$ themselves do not commute. Instead, applying the nowhere-vanishing transformation
\begin{equation}\label{nonvanishing}
    \Psi \rightarrow e^{-\pi \phi y^2} \Psi = \tilde \Psi
\end{equation}
to all states in the Hilbert space, and the corresponding transformation 
\begin{equation}
    \mathcal{O} \rightarrow e^{-\pi \phi y^2} \mathcal{O} e^{\pi \phi y^2} =  \tilde{\mathcal{O}} 
\end{equation}
to the operator algebra, the raising operator becomes ${\hat a}^\dagger = \partial_z$, so that wavefunctions in higher Landau levels are derivatives of the LLL wavefunctions. Finally, applying the inverse of transformation (\ref{nonvanishing}), we find
\begin{equation}\label{LLfuncs}
    \Psi_{a,b}^{(j)}(x,y) = \frac{1}{m^j} \sum_{k=0}^j \binom{j}{k} \of{i\sqrt{\frac{\pi \phi m^2}{2}}}^k \frac{\partial^{j-k}\vartheta_{a,b}}{\partial z^{j-k}}\of{\frac{z}{m}} H_k(\sqrt{2\pi \phi} y)  e^{-\pi \phi y^2},
\end{equation}
where $H_k$ denotes the $k$-th Hermite polynomial. Note that we have not normalized our wavefunctions, as the present, un-normalized form is more conducive to the construction of a local hopping Hamiltonian with the desired features, as we discuss in the following section.

We further note that on a lattice, magnetic translational symmetry with a
{\em finite} magnetic unit-cell (i.e. commensurate flux) can be recast as
ordinary translational symmetry via a ``singular gauge transformation''
\cite{franz_quasiparticles_2000,vafek_quasiparticles_2001,franz_qed_2002}. This observation highlights the close relationship
between Hofstadter-type constructions and Chern-insulator models, and may
prove useful when applying our framework in a broader FCI context. Since it
does not simplify the analysis pursued here, we will not make use of this
transformation in the following.
We now proceed to construct the lattice
Hamiltonians realizing the desired flat-band structure.

\section{A lattice model with Landau-level-like flat bands\label{onebod}}

The continuum torus Landau-level wavefunctions \eqref{LLfuncs}, having good quantum numbers under the translations $T_x^m$ and $T_y^n$, immediately give rise to magnetic Bloch states on any lattice (on the same torus) whose magnetic unit-cell is spanned by the associated translation vectors. Their significance to our problem  is the following: For a given number $p$, one may identify a Landau-level filling-factor $\nu$ such that a delta-function contact interaction has a unique bosonic zero energy wavefunction constructed from the first $p$ Landau levels $\Psi^{(j)}_{a,b}$, $j=0\dots p-1$ up to topological degeneracy. The same wavefunction, at the corresponding lattice filling-factor, is then also a zero-energy state of contact interactions when restricted to the lattice, and will, moreover, be an eigenstate of a suitable one-body Hamiltonian if the (lattice restricted) $\Psi^{(j)}_{a,b}$ for $j=0\dots p-1$ form degenerate flat bands as $a$, $b$ range over the Brillouin zone. Furthermore, it will be a unique ground state (again  modulo topological degeneracy) under the additional mild assumptions that all other bands are separated from the flat bands by an energy gap, and that the magnetic unit-cell has ``enough'' sites such that one is sufficiently close to the continuum limit. How many sites are required to preserve the ground state uniqueness of  the continuum model is left to case-by-case study, and we will present numerics for some cases of interest. In this section, we will construct an exponentially decaying positive semi-definite hopping Hamiltonian with
$p$ zero-energy flat bands 
given by the lattice restrictions of the $\Psi^{(j)}_{a,b}$, for $j=0\dots p-1$.


We will now write $mn=p+q$ for the number of sites per unit cell, where $p$ is the number of zero-energy bands, and $q$ the number of finite energy dispersing bands.
We begin with the case $q=1$ for clarity. Let $\Phi(x,y)$ be a first-quantized (continuum) wavefunction. Restricting
it to the lattice sites $(x_\alpha,y_\alpha)$ of the $mn$-site magnetic
unit-cell defines a vector with components
\begin{equation}\label{genvec}
    \phi_\alpha = \Phi(x_\alpha,y_\alpha),
\end{equation}
where $\alpha$ indexes the sites of the unit cell in some fixed order.
Thus each continuum wavefunction defines a vector in the magnetic
unit-cell Hilbert space $\mathcal H \cong \mathbb{C}^{mn}$.
Similarly, define $\phi^\alpha = {\of{\phi_\alpha}}^\dagger$.
In the following, we will use the Einstein summation convention for the vectors in $\mathcal{H}$, i.e. 
\begin{equation}
    \psi^\alpha \phi_\alpha = \sum_{\alpha \in \text{U.C.}} \braket{\psi|x_\alpha,y_\alpha}\braket{x_\alpha,y_\alpha|\phi}. 
\end{equation}

Using the vectors induced by the functions $\Psi_{a,b}^{(j)}(z)$, we define 
\begin{equation}\label{q=1}
    \of{\Psi^\perp_{a,b}}_\alpha = \epsilon_{\alpha\alpha_0\dots\alpha_{mn-2}}\of{\Psi^{(0)}_{a,b}}^{\alpha_0}\dots \of{\Psi^{(mn-2)}_{a,b}}^{\alpha_{mn-2}},
\end{equation}
which we extend to a state $\ket{\Psi^\perp_{a,b}}$ defined on the entire lattice by demanding that it be an eigenstate of $T_x^m$ and $T_y^n$ with the appropriate eigenvalue, so that in first quantization
\begin{align*}
    \Psi_{a,b}^\perp(x+Am,y+Bn)  &= e^{i2\pi(aA-bB)}e^{-i2\pi\phi Bnx}\Psi^\perp_{a,b}(x,y). 
\end{align*}
We caution the reader that, unlike the $\Psi_{a,b}^{(j)}(x,y)$, the functions $\Psi^\perp_{a,b}(x,y)$ as presented are not defined away from the lattice points of the full torus. With the $\ket{\Psi_{a,b}^\perp}$ determined as states in the full Hilbert space, we may define the one-body Hamiltonian 
\begin{equation}
    H = \sum_{a,b} \ket{\Psi_{a,b}^\perp} \bra{\Psi_{a,b}^\perp},
\end{equation}
or in second quantization
\begin{align}
    H = \sum_{R,R'} t(R,R') c^\dagger_R c_{R'}
\end{align}
with 
\begin{align}
    t(R,R') &= \sum_{a,b} \braket{R|\Psi_{a,b}^\perp}\braket{\Psi_{a,b}^\perp|R'} \nonumber \\
        &= e^{-i2\pi \phi (B-B')nx} \sum_{a,b}e^{i2\pi (a(A-A') - b(B-B'))} \nonumber \\
        & \ \ \times \epsilon_{r \alpha_0\dots \alpha_{mn-2}} \Psi^{(0)}_{a,b}\of{x_{\alpha_0},y_{\alpha_0}}\dots \Psi^{(mn-2)}_{a,b}\of{x_{\alpha_{mn-2}},y_{\alpha_{mn-2}}} \nonumber \\
        & \ \ \times \epsilon^{r' \alpha'_0\dots \alpha'_{mn-2}} \overline{\Psi^{(0)}_{a,b}\of{x_{\alpha'_0},y_{\alpha'_0}}}\dots \overline{\Psi^{(mn-2)}_{a,b}\of{x_{\alpha'_{mn-2}},y_{\alpha'_{mn-2}}}}.
\end{align}
By construction, $H$ annihilates $\ket{\Psi^{(j)}_{a,b}}$ for all $a$, $b$, and $0 \leq j < mn-1$, and by the unitarity of the magnetic translation operators
\begin{equation}
    \braket{\Psi_{a,b}^\perp|\Psi_{a',b'}^\perp} = \delta_{a,a'}\delta_{b,b'}\braket{\Psi_{a,b}^\perp|\Psi_{a,b}^\perp}, 
\end{equation}
meaning 
\begin{equation}
    H\ket{\Psi_{a,b}^\perp} = \ket{\Psi_{a,b}^\perp}\braket{\Psi_{a,b}^\perp|\Psi_{a,b}^\perp}.
\end{equation}
It follows that our model is gapped so long as $\braket{\Psi_{a,b}^\perp|\Psi_{a,b}^\perp} > 0$, which by inspection of (\ref{q=1}) is equivalent to the statement that the vectors $\{\ket{\Psi^{(j)}_{a,b}}\}_{j=0,\dots,mn-2}$ are linearly independent. Though this fact is not guaranteed \textit{a priori}, the orthogonality of the continuum wavefunctions $\Psi^{(j)}_{a,b}(x,y)$ implies that the vectors $\ket{\Psi^{(j)}_{a,b}}$ will become linearly independent almost certainly for large enough $mn$. Furthermore, we remark that the precise Cartesian coordinates
$(x_\alpha, y_\alpha)$ of the sites in the magnetic unit cell may be treated
as variables of our model and can therefore be used to lift any accidental
linear dependence among the $\ket{\Psi^{(j)}_{a,b}}$. In principle, the sites
within the magnetic unit cell need not be related by primitive lattice
translations, i.e., it is not required that
$(x_{\alpha}-x_{\alpha'}, y_{\alpha}-y_{\alpha'}) \in \mathbb{Z}^2$.
For simplicity, however, we choose a regular lattice basis.
In practice, we found the choice
$x_\alpha = \alpha_x$, $y_\alpha = \alpha_y + \pi/2n$ for
$\alpha = \alpha_x + n\alpha_y$ with
$0 \le \alpha_x < m$, $0 \le \alpha_y < n$
to work particularly well for maximizing the Gram determinant of the 
$(\Psi^{(j)}_{a,b})_\alpha$ vectors.

We therefore proceed under the assumption that the
$\ket{\Psi^{(j)}_{a,b}}$ are linearly independent. Finally, we normalize $H$
so that the natural energy scale given by the real-space amplitude
$t_0 = \max_{R}|\braket{R|H|R}|$ is equal to one.

It is non-trivial to establish analytically that the Hamiltonian constructed above is exponentially local in the sense that
\[
|\langle R|H|R'\rangle| \le \exp(-|R-R'|/\xi)
\]
as $|R-R'|\to\infty$ for some finite length scale $\xi$. The crucial point, however, is that although $H$ is assembled from
rank-one operators associated with magnetic Bloch states, it is
\emph{not} itself a spectral projector onto a flat band. In particular, we deliberately refrain from normalizing the individual $\ket{\cdot}\!\bra{\cdot}$ terms. While such a normalization would be natural if the sole goal were to construct an exact projector, it would simultaneously enforce flattening of {\em all} bands and disrupt the analytic structure underlying locality of the real space Hamiltonian.

Instead, by retaining the unnormalized form, the resulting Hamiltonian acquires a dispersive finite-energy sector while maintaining an exactly flat zero-energy band. This dispersion plays a central role: it allows the matrix elements of $H$
to be expressed as analytic functions of the magnetic Bloch data, whose
Fourier coefficients decay exponentially in real space, with a characteristic
length set by the imaginary gap of the analytic continuation in momentum
space. This mechanism is closely related to the analyticity--locality correspondence discussed in Ref.~\cite{lee_band_2016}, where it is emphasized that exact band-projection (onto Chern bands) generically leads to long-range hopping.
In contrast, retaining dispersion preserves the analytic structure required for locality.

Although we do not attempt a rigorous proof of exponential locality for this class of Hamiltonians, clear numerical evidence for exponentially decaying hopping amplitudes, independent of system size, is observed in all examples studied: see Figs.~\ref{hoppingParameter}, ~\ref{1DhoppingParameter} and ~\ref{hoppingParameter3D} below. 


It is desirable to generalize these results to the case $q>1$.
This allows the continuum limit to be approached in a controlled way:
smaller flux per plaquette corresponds to a larger magnetic unit-cell 
size $mn$, while keeping the number of flat bands $p$ fixed. Since the
total number of bands is $mn$, increasing the unit-cell size requires an
increase in the number of dispersing bands,
$q = mn - p$. The $q>1$ case can be realized by building on the
$q=1$ construction.
As before, we encode orthogonality with the Levi-Civita symbol, $\of{\Psi^\perp_{a,b}}_\alpha \propto \epsilon_{\alpha\alpha_0\dots\alpha_{mn-2}}\of{\Psi^{(0)}_{a,b}}^{\alpha_0}\dots \of{\Psi^{(p-1)}_{a,b}}^{\alpha_{p-1}}$, but now after contraction $mn-p = q > 1$ indices remain so that this procedure alone does not specify a vector. To handle the remaining indices, we supply $q-1$ vectors from a basis $\{\ket{\Phi^\beta}\}$ of the magnetic unit-cell Hilbert space $\mathcal{H}$. If one wishes, the chosen basis can be thought of as a free parameter of the model, though in practice it is easiest to work with the natural basis $\{\ket{e^\beta}\}$ of Kronecker-delta wavefunctions supported on single lattice sites labeled by $\beta$, for which $\ket{\phi} = \phi_\alpha \ket{e^\alpha}$: note that we implicitly introduced this basis in (\ref{genvec}). 
We then define
\begin{align}\label{psiperpbeta}
    \of{\Psi_{a,b}^\perp(\vec\beta)}_\alpha &= \epsilon_{\alpha \alpha_0\dots\alpha_{mn-2}} \of{\Psi_{a,b}^{(0)}}^{\alpha_0}\dots \of{\Psi_{a,b}^{(p-1)}}^{\alpha_{p-1}} \nonumber \\
        & \ \ \times \of{\Phi^{\beta_1}}^{\alpha_{p}} \dots \of{\Phi^{\beta_{q-1}}}^{\alpha_{mn-2}},
\end{align}
where $\vec\beta = (\beta_1,\dots\beta_{q-1})$, which, for $\ket{\Phi^\beta} = \ket{e^\beta}$, simplifies to
\begin{equation}
    \of{\Psi_{a,b}^\perp(\vec\beta)}_\alpha = \epsilon_{\beta_1\dots \beta_{q-1}\alpha \alpha_0\dots \alpha_{p-1} } \of{\Psi_{a,b}^{(0)}}^{\alpha_0}\dots \of{\Psi_{a,b}^{(p-1)}}^{\alpha_{p-1}}.
\end{equation}
Extending to the full lattice in the usual way, we write
\begin{equation}\label{H_one_body}
    H = \sum_{a,b} H_{a,b} = \sum_{a,b}\sum_{\vec\beta} C_{\vec\beta} \ket{\Psi_{a,b}^\perp(\vec\beta)} \bra{\Psi_{a,b}^\perp(\vec\beta)},
\end{equation}
where the coefficients $C_{\vec\beta}$ are free parameters of the model that control the behavior of the non-flat bands. Again, in second quantization we have
\begin{align}
    H = \sum_{R,R'} t(R,R') c^\dagger_R c_{R'}
\end{align}
with 
\begin{align}
    t(R,R') &= \sum_{a,b}\sum_{\vec \beta} C_{\vec\beta}\braket{R|\Psi_{a,b}^\perp}\braket{\Psi_{a,b}^\perp|R'}\\
        &= e^{-i2\pi \phi (B-B')nx} \sum_{a,b}e^{i2\pi (a(A-A') - b(B-B'))} \nonumber \\
        & \ \ \times \sum_{\vec \beta} C_{\vec\beta} \ \epsilon_{r \alpha_0\dots \alpha_{p-1} \beta_1\dots \beta_{q-1}} \of{\Psi_{a,b}^{(0)}}^{\alpha_0}\dots \of{\Psi_{a,b}^{(p-1)}}^{\alpha_{p-1}} \nonumber \\
        & \ \ \times\epsilon^{r' \alpha'_0\dots \alpha'_{p-1} \beta_1\dots \beta_{q-1}} \of{\Psi_{a,b}^{(0)}}_{\alpha'_0}\dots \of{\Psi_{a,b}^{(p-1)}}_{\alpha'_{p-1}}.
\end{align}
We will focus on the simplest case $C_{\vec\beta}\equiv 1$, for which one can show that
\begin{equation}
H=\sum_{a,b}\mathcal{E}_{a,b}\,P^\perp_{a,b},\label{projsum}
\end{equation}
where $\mathcal{E}_{a,b}=\det_{ij}\!\bigl(\langle\Psi^{(i)}_{a,b}|\Psi^{(j)}_{a,b}\rangle\bigr)$ is the Gram determinant, i.e.\ the volume of the parallelepiped spanned by the lattice-restricted Landau-level wavefunctions, and $P^\perp_{a,b}$ denotes the orthogonal projection onto $\bigl(\mathcal{H}^{\parallel}_{a,b}\bigr)^\perp=\bigl(\mathrm{span}\{\Psi^{(0)}_{a,b},\dots,\Psi^{(p-1)}_{a,b}\}\bigr)^\perp$. The details of this derivation are presented in Appendix~\ref{appA}.

Eq.~\eqref{projsum} exhibits several notable structural features. First, it is
independent of the choice of basis $\{\ket{\Phi^\beta}\}$ (Appendix~\ref{appA}).
Second, the states $\ket{\Psi^{(j)}_{a,b}}$, $j=0,\dotsc,p-1$, span a
zero-energy subspace of $H$, while the remaining $q=mn-p$ bands are automatically
degenerate with one another and share the same dispersion relation
$\mathcal{E}_{a,b}$. As before, a linear dependence among the Bloch states
$\ket{\Psi^{(j)}_{a,b}}$ would signal a band touching; however, this does not
occur in the cases studied numerically, as demonstrated in Fig.~\ref{cutoffBS}
below.

Most importantly for our purposes, the above shows that the
mechanism underlying locality is unchanged from the $q=1$ case. Since
the Hamiltonian is again assembled from unnormalized rank-one data, the
analytic argument establishing exponential localization carries over
directly to the general $q>1$ construction. Figs.~\ref{hoppingParameter}, ~\ref{1DhoppingParameter} and ~\ref{hoppingParameter3D} further demonstrate for representative
$q>1$ realizations that the hopping amplitudes remain exponentially
localized. In the following section we examine these same cases in the
presence of interactions.


\section{The interacting model: parent Hamiltonians for lattice parton states\label{results}}
In this section we apply the general construction to several
representative cases. We begin with the Abelian Jain $21$-state and then
consider the non-Abelian $22$- and $33$-states, which support Ising- and
Fibonacci-type anyons, respectively. For each case we construct the
corresponding lattice Hamiltonian and confirm the expected zero-mode
structure through exact diagonalization of small systems.

\subsection{General theoretical framework}

\begin{figure}[ht]
    \centering
    \includegraphics[width=\linewidth]{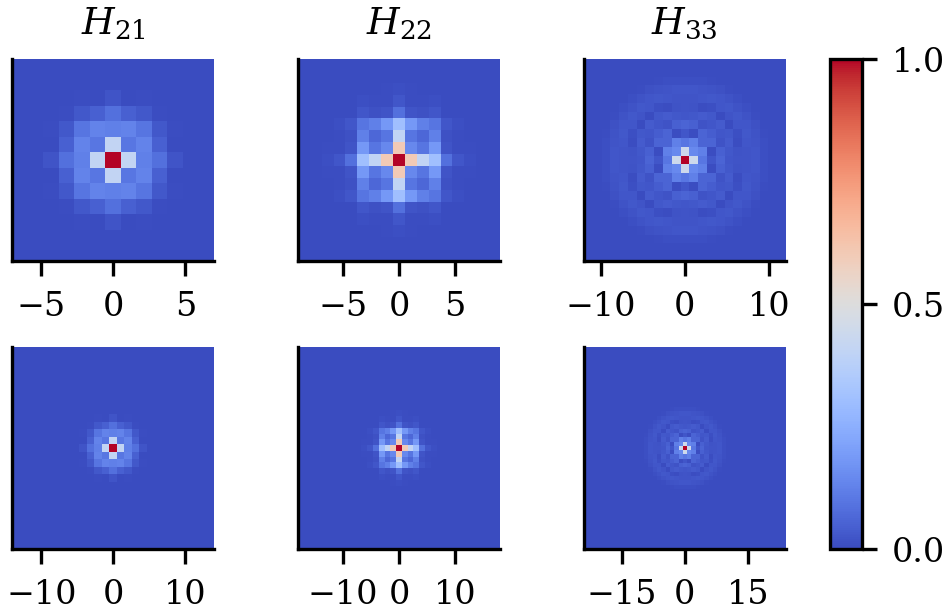}
    \caption{A plot of the hopping parameter $t_R = |\braket{R_0|H|R}|$ for the 21-, 22-, and 33-state parent Hamiltonians at two different system sizes. Plots in each column are computed with a four, four and 12-site magnetic unit-cell for $H_{21}$, $H_{22}$ and $H_{33}$ respectively. In particular, the small lattices in the top row are of sizes $14\times14$, $18 \times 18$ and $24 \times 24$ again for the 21-, 22-, and 33-state parent Hamiltonians respectively, while the large lattices in the lower row have dimensions $28\times 28$, $36 \times 36$ and $48 \times 48$. Unless explicitly stated otherwise, all Hamiltonians here and elsewhere are normalized so that $t_0 = 1$. Distances are given in terms of the lattice constants $a = a_x = a_y$.  \label{hoppingParameter}}
\end{figure}
\begin{figure}[ht]
    \centering
    \includegraphics[width=\linewidth]{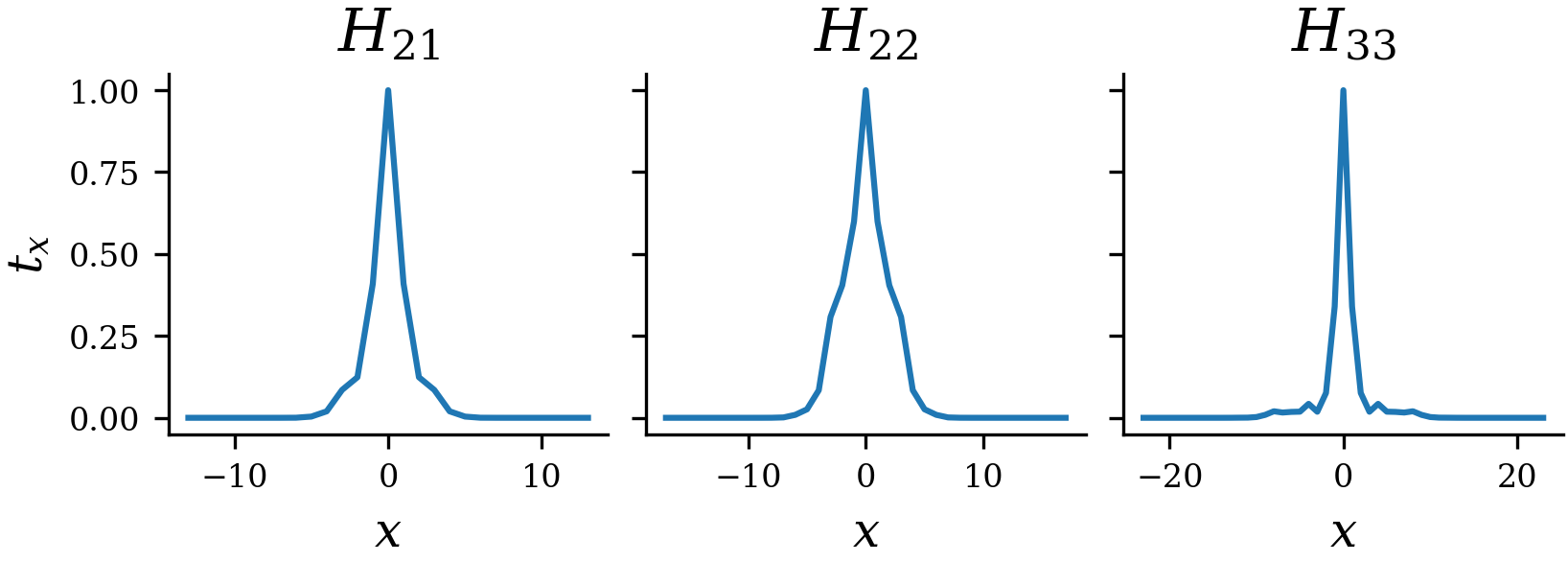}
    \caption{A one-dimensional slice of the hopping parameter plot in Fig.~\ref{hoppingParameter} i.e. for $t_x = |\braket{R_0|H|x,0}|$ for the 21-, 22-, and 33-state parent Hamiltonians, clearly demonstrating exponential decay of the hopping strength.   \label{1DhoppingParameter}}
\end{figure}
\begin{figure}
    \centering
    \includegraphics[width=\linewidth]{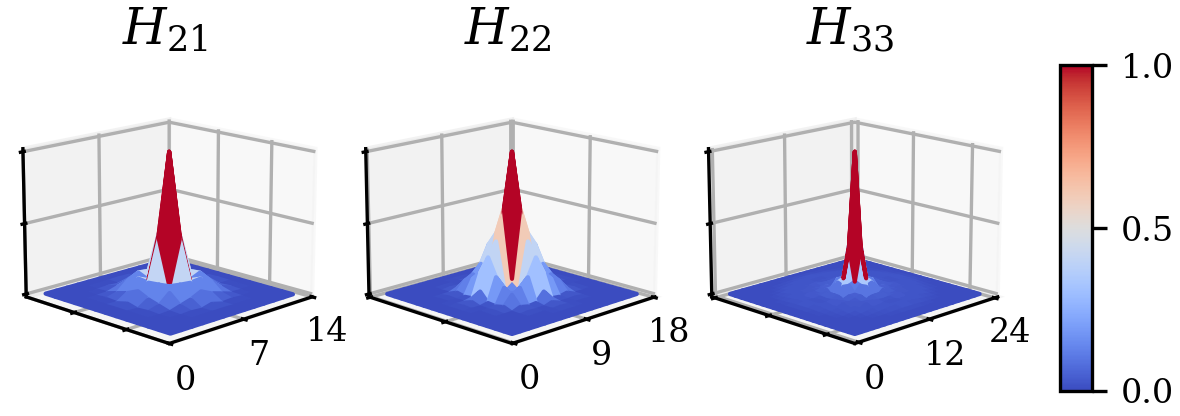}
    \caption{A three-dimensional plot of the exponentially-decaying hopping parameter $t_R = |\braket{R_0|H|R}|$ for the 21-, 22-, and 33-state parent Hamiltonians. Plots for $H_{21}$ and $H_{22}$ are generated for lattices of size $14 \times 14$ and $18 \times 18$ with four site magnetic unit-cells, while the plot for $H_{33}$ corresponds to a $28 \times 28$ site lattice and a 12-site magnetic unit-cell. \label{hoppingParameter3D}}
\end{figure}
 \begin{figure}[ht]
    \centering
    \includegraphics[width=\linewidth]{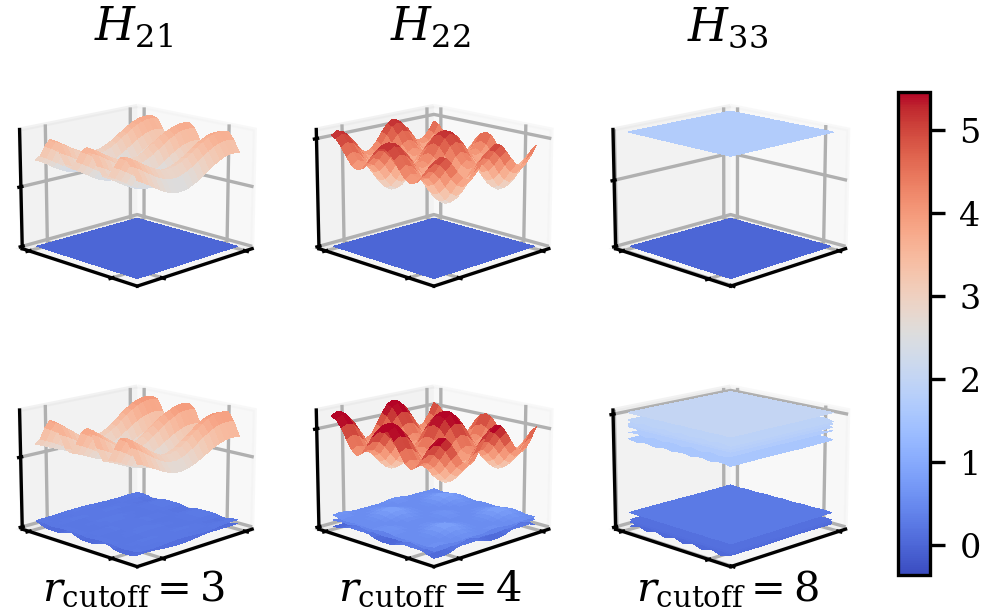}
    \caption{Band structure plots with and without finite-distance cutoffs, see (\ref{rcoeq}), for the three parent Hamiltonians discussed in the main text. Plots are computed for lattices of size $60 \times 20$, $40 \times 40$ and $24 \times 24$ with three, four and 12-site magnetic unit-cells for $H_{21}$, $H_{22}$ and $H_{33}$ respectively.  \label{cutoffBS}}
\end{figure}

The Hamiltonian (\ref{H_one_body}) can be promoted to an FCI Hamiltonian by choosing an appropriate interaction term. To make contact with solvable FQH parent Hamiltonians in the continuum, we implement the strategy described initially, and switch on positive contact-interactions for bosonic particles. Computationally, the simplest choice is to work with hard-core Bosons, thus to take the interaction strength to infinity, as this reduces the size of the Hilbert space.
 As established above, any zero mode of the bosonic continuum model with
$p$ degenerate Landau levels on a torus with $L=MN\phi$ flux quanta in the
presence of $\delta^2(z_i-z_j)$ contact interactions induces a
corresponding zero mode on the lattice. One therefore expects the
number of zero modes to coincide with that of the continuum model. There are two effects associated with our lattice discretization that could change this outcome. First, the condition for linear dependence is weaker when evaluated on the lattice, as the set of particle configurations is discrete and finite. Thus, zero modes that are linearly independent in the continuum could become linearly dependent on the lattice, lowering the number of zero modes. Similarly the condition for having zero interaction-energy is weaker as well. Therefore, there could be zero modes on the lattice that have no
correspondence in the continuum model, increasing the total number of
zero modes. Both effects are expected to disappear for sufficiently
large magnetic unit-cells, or equivalently for large $q=mn-p$, where the
relevant conditions approach their continuum limits.

In this section, we will numerically study whether given instances of our lattice model faithfully reproduce the zero-mode space of the continuum model. Where confirmed, this will be strong evidence that the universality class of the continuum model survives our lattice discretization. The most direct evidence will come from the zero-mode degeneracy at the incompressible filling-factor, which on the torus determines the number of anyon types \cite{simon_topological_2023}. More broadly, zero-mode counting in the presence of additional flux quanta, or at lower particle density, encodes information about fusion rules and is deeply related to the edge conformal field theory. 

To this end, we performed numerics for small system-sizes for the lattice model only. For the continuum side, we can rely on a combination of field theoretic methods and recent results on the relevant continuum models, as we will now explain. Consider the $rr$-bosonic parton state 
\begin{equation}\label{psi_rr}
\psi_{rr}(z,\bar z)=\chi_r(z,\bar z)^2.
\end{equation}
Here, $(z,\bar z)$ is short for the collection of complex coordinates $(z_1\dotsc z_N,\bar z_1\dotsc \bar z_N)$, and $\chi_r$ is the wavefunction of a $\nu=r$ integer quantum Hall state. In the following we elide the obligatory Gaussian factors. The state $\psi_{rr}$ satisfies an $\mathcal{M}=2$ clustering property in that it has a second-order zero when the coordinates of one particle approach those of another.
It is natural to conjecture that $\psi_{rr}$ is the unique densest state with this property within the lowest $2r-1$ Landau levels. This would then render the state the unique densest zero mode for contact interactions of the form $\delta^2(z_i-z_j)$, projected onto $r$ Landau levels. Properties of this kind were first conjectured by Wen \cite{wen_non-abelian_1991}, who also observed that giving proofs for such claims is highly non-trivial. This may be understood when comparing the situation for $r>1$ to that for $r=1$ (LLL), where statements reduce to facts about symmetric polynomials that can be studied with traditional methods of commutative algebra \cite{stone_schur_1990}. These methods do not generalize to the situation of non-holomorphic polynomials with $z$ and $\bar z$ present, as in Eq. \eqref{psi_rr}, that are furthermore subject to constraints on powers of $\bar z$ (due to LL projection). Only quite recently were methods developed to deal with this problem rigorously \cite{bandyopadhyay_entangled_2018}, originally for the fermionic 221-state, and subsequently the 222-state \cite{tanhayi_ahari_partons_2023}. Reference \cite{tanhayi_ahari_partons_2023} also discusses some features of the general situation, though focusing on Fermions. Most results can be easily adopted to Bosons, as we will explain. 

The general approach of
Refs.~\cite{bandyopadhyay_entangled_2018,bandyopadhyay_local_2020,
tanhayi_ahari_partons_2023,cruise_sequencing_2023}
establishes rigorous results concerning parton clustering properties and
the associated zero-mode solution spaces only for the disk and other
zero-genus geometries. Nevertheless, these methods provide considerable
insight into the corresponding situation on the torus relevant to the
present work. We will therefore briefly describe the most important features of these methods. 
We may decompose the $N_e$-particle state $\psi_{rr}$ into occupancy eigenstates labeled by partitions
$\{\ell_i\}_{i=1\dotsc N_e}$, where $\ell_i$ is the angular momentum of the $i$th particle, such that $\ell_i\leq\ell_{i+1}$, and $J=\sum_i\ell_i$ is the total angular momentum of the state. One may also pass to an occupation-number representation $\{n_\ell\}_{\ell=0,1,\dots}$ where $n_\ell$ is the number of $\ell_i$ equal to $\ell$. A partition $\{\ell_i\}$ is said to dominate another partition $\{\ell'_i\}$ if $\sum_{i=1}^k (\ell_i - \ell'_i) \geq 0 $ for all $k = 1, \dots, N_e$, and we write $\{\ell_i\} \succeq \{\ell'_i\}$; note that $\succeq$ only defines a partial ordering as it can happen that neither $\{\ell_i\} \succeq \{\ell'_i\}$ nor $\{\ell'_i\} \succeq \{\ell_i\}$ obtains. Nevertheless, among the partitions occurring in the spectral decomposition of the state $\psi_{rr}$, there is a ``root partition'' that dominates all other partitions in this decomposition, $\{\ell_i^{\sf root}\}\succeq \{\ell_i\}$. For a product wavefunction $\psi=\psi_1\psi_2$, with factors having unique known root partitions $\{\ell_i^{1,2}\}$, the unique root partition is given by $\ell_i=\ell^1_i+\ell^2_i$. In the case of $\psi_{rr}$, each factor $\chi_r$ consists of a single Slater determinant that is its own root partition, whose occupancy pattern is trivially given by $\{n_\ell\}=\{rrrrr\dotsc\}$, or $\ell_i=\lfloor(i-1)/r\rfloor$. The above rules result in the root pattern $\{n_\ell\}=\{r0r0r0r\dotsc\}$ for $\psi_{rr}$. One immediately infers the filling factor $\nu=r/2$ of the state.

In the lowest Landau-level, it has long been known that root patterns of polynomials that are solutions to certain clustering properties are subject to certain local constraints \cite{feigin_differential_2002, bernevig_model_2008, bernevig_generalized_2008} called ``generalized Pauli principles'' (GPPs). 
For mixed Landau-level situations like the present one, or for general
multi-component states, the root pattern does not correspond to a single
Fock basis state. One must therefore distinguish the {\em root pattern} from
the {\em root state}, defined as the projection of the wavefunction onto all
Fock basis states consistent with the root pattern. These root states can, in general, be entangled, owing to different configurations of Landau-level, layer, or valley degrees of freedom present. We will generically refer to these degrees of freedom as ``pseudospin'', with a pseudospin $S=r-1$ signifying that the $rr$-state lives in $2r-1$ Landau levels. It has been observed recently  that the root states of zero modes of general frustration free two-body \cite{ bandyopadhyay_entangled_2018,bandyopadhyay_local_2020, tanhayi_ahari_partons_2023} or $k$-body \cite{ cruise_sequencing_2023} interactions, e.g., interactions enforcing clustering properties, are still characterized by local constraints termed ``entangled Pauli principles'' (EPPs).
The rigorous identification of the $rr$-state as the unique densest
zero mode of its parent Hamiltonian then follows from the framework
developed in these works: one establishes an EPP for the problem and
shows that the root state of $\psi_{rr}$ is the unique densest
configuration consistent with it, which in turn implies that
$\psi_{rr}$ is the unique densest zero mode.

The EPP-toolkit provides ways to do rigorous zero-mode counting for frustration-free parent Hamiltonians and related constraint problems in multi-component/multi-LL settings in genus-zero geometries. It does not offer rigorous insight into the same problem on the torus. However, root states are manifestly projected out when the thin-cylinder limit is taken, which is locally the same as the thin torus limit.  Indeed, it is well studied in the lowest Landau level that the 
thin-torus limit is adiabatically connected to tori of any finite aspect ratio \cite{Seidel_Incompressible_2005, seidel_abelian_2006}, and that the counting induced by the GPP, together with periodic boundary-conditions, gives the correct zero-mode counting on the torus as well \cite{seidel_gapless_2011}. 
We will assume that this is also true in situations governed by EPPs, and the numerical results discussed below will lend additional support to this assumption.

As a first step toward counting zero modes on the torus for the models
considered here, we revisit the state $\psi_{rr}$ of Eq.~\eqref{psi_rr}
in the simpler setting of zero-genus geometries. We modify one of the factors $\chi_r$ by introducing holes at the beginning, such that the new root pattern of this factor is $\{(r-k)rrrrrr\dotsc\}$.  The product is then still a zero mode of the same contact Hamiltonian, satisfying the same clustering condition. Its root pattern is $\{r-k,k,r-k,k,r-k,\dotsc\}$. Specifically, in disk
geometry this configuration increases the angular momentum of the
state, which can be pictured as resulting from an outward charge pump
that leaves behind quasihole excitations near the origin while
preserving the bulk filling factor. On the torus, however, the
underlying root state can be made compatible with periodic boundary
conditions, so that it satisfies the same EPP when formulated with
periodic boundary conditions. These states must then be interpreted as ``thin-torus'' zero-mode states of uniform density, suggesting that the torus degeneracy of the $rr$-state is $r+1$.
This is absolutely consistent with the field-theoretic ``parton logic'', which features an internal 
$SU(2)$ gauge structure and yields an 
$SU(2)_r$ non-Abelian sector coupled to a 
$U(1)$ charge sector after enforcing the gauge constraint \cite{wen_projective_1999,anand_realspace_2022}.

In the following, we will also be interested in parton states of the form
\begin{equation}\label{psi_rr_1}
\psi_{r.r-1}(z,\bar z)=\chi_r(z,\bar z)\chi_{r-1}(z,\bar z).
\end{equation}
The two IQH Slater-determinant factors of the terms on the RHS have root patterns $\{rrr\dotsc\}$ and $\{r-1,r-1,r-1\dotsc\}$. The associated root partitions $\{\ell^{1,2}_i\}$ have incommensurate strides $r$ and $r-1$, respectively. Via the rules discussed above, this leads to a root pattern (somewhat complicated for general $r$) for $\psi_{r,r-1}$ with a unit cell containing $r(r-1)$ particles, spanning $2r-1$ sites.
Therefore, the state has filling factor $\nu=r(r-1)/(2r-1)=(1/r+1/(r-1))^{-1}$.
For example, below we will explicitly consider $r=2$, for which we obtain the root pattern $110110\dotsc$ at $\nu=2/3$.
Using the same logic described for the $rr$-state, we may conclude that the $r(r-1)$-state \eqref{psi_rr_1} is the unique densest state with the $\mathcal{M}=2$ clustering property (zero mode of the $\delta^2(z_i-z_j)$ interaction) in $2r-2$ Landau levels.

We may moreover examine the torus degeneracy of the state using the same method described above. To this end, we look at the root pattern of the $r(r-1)$-state in the presence of holes introduced at the beginning of one (or both) of the Slater-determinants. Up to boundary effects, this only shifts the root pattern we had obtained in the absence of such holes. This, again, is a result of the incommensurability between the root partitions of the two factors. The interpretation of this is that all torus ground states are related by translation, where the torus degeneracy is the size of the unit cell of the root pattern, or $2r-1$. This ``minimum torus degeneracy'' (dictated by magnetic translations alone) signifies an Abelian state. This conclusion is again consistent with the usual parton effective-field-theory logic: treating the two factors as integer quantum-Hall states with unequal Chern numbers and enforcing the internal gauge constraint yields a multi-component Abelian Chern–Simons theory, reproducing the filling factor $\nu=r(r-1)/(2r-1)$ and a ground-state degeneracy $2r-1$.
 
\begin{figure}
    \centering
    \includegraphics[width=0.75\linewidth]{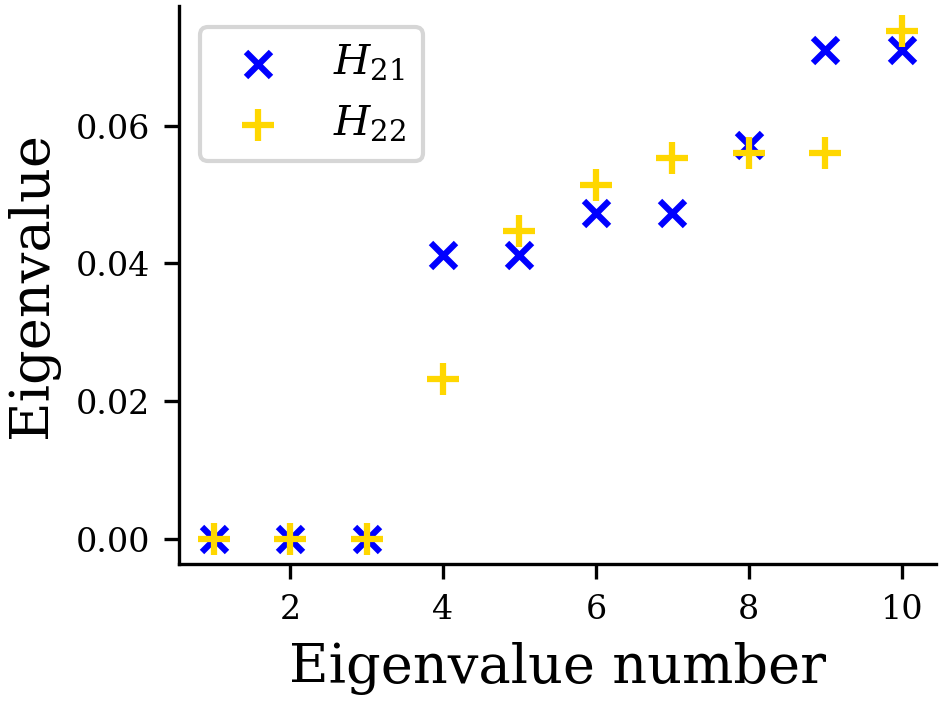}
    \caption{The ten smallest eigenvalues of the 21- and 22-state parent Hamiltonians. The eigenvalues of both $H_{21}$ and $H_{22}$ are computed for four particles on a $4 \times 6$-site lattice with unit cells of dimension $m = n = 2$. The three smallest eigenvalues in each case are zero up to finite machine precision, in agreement with the expected topological ground state degeneracy.\label{gsspec}}
\end{figure}

\subsection{The 21-state -- Abelian bosonic composite fermions}

We begin by presenting numerics 
on realizations of our model stabilizing the
Jain-$21$ state, i.e. the $r=2$ case of Eq. \eqref{psi_rr_1}, which is a bosonic composite Fermion state. Its 2LL-projected parent Hamiltonian in the continuum is one of a well-established family \cite{bandyopadhyay_local_2020}. 
Earlier literature on Jain composite-Fermion parent Hamiltonians focused on the fermionic Jain-$211$ (filling factor $\nu=2/5)$ state \cite{rezayi_origin_1991,chen_jain_2017, Chen_construction_2020} and the zero modes of its parent Hamiltonian, which differ (in disk geometry) from those of the Jain-$21$ parent Hamiltonian precisely by a single Jastrow factor $\prod_{i<j}(z_i-z_j)$.
Hence, at least in zero-genus geometries,
the zero modes of the continuum problem are well-established, with the Jain-$21$ state being the unique densest zero mode at filling factor $\nu=2/3$.


Since $\psi_{21}$ occupies two Landau levels, we require $p=2$ flat bands. To this end, we realize our model with a magnetic unit-cell of four sites, which yields two flat bands and two additional dispersing bands. We denote instances of the model with two flat bands by $H_{21}$, indicating that they serve as parent Hamiltonians of the lattice $21$-state. In principle, an arbitrary number of dispersing finite-energy bands may
be present; unless otherwise stated, however, the realizations of
$H_{21}$ considered here contain two such bands, corresponding to a
magnetic unit-cell with $m=n=2$.
In Figs.~\ref{hoppingParameter}, ~\ref{1DhoppingParameter} and \ref{hoppingParameter3D}, we present the single-particle hopping amplitude $t_R = |\braket{R_0|H_{21}|R}|$ for $R_0 = (0,0)$ computed on two lattices of sizes $14 \times 14$ and $28 \times 28$; here, and for all other models examined, we find that the single-particle hopping amplitude has no dependence on system size once the lattice is sufficiently large, and that the data are consistent with exponential localization, as anticipated in Sec.~\ref{onebod}. We also give plots of the band structure in Fig.~\ref{cutoffBS} for a lattice of size $60 \times 20$ with a three-site magnetic unit-cell. 

To examine the many-body spectrum, we choose lattice dimensions
$4\times6$, so that each band contains six single-particle states,
corresponding to four particles at filling $\nu=2/3$. For this Abelian FQH state, one expects the minimal torus degeneracy of three — equal to the denominator of the filling fraction \cite{oshikawa_fractionalization_2006} — consistent with the magnetic translation group and the root pattern $110110110\dots$ \cite{bandyopadhyay_local_2020}. In Fig.~\ref{gsspec}, we show the lowest ten eigenvalues of $H_{21}$ for this system. The expected threefold ground-state degeneracy is clearly visible.



\begin{figure}[ht]
    \centering
    \includegraphics[width=0.75\linewidth]{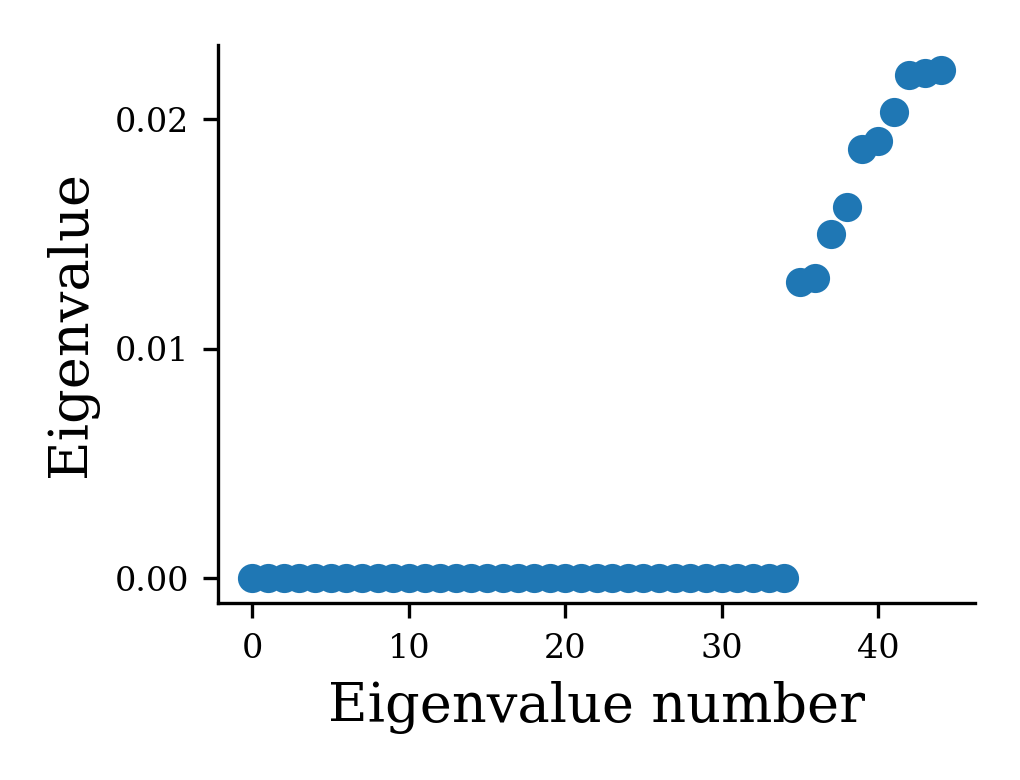}
    \caption{Plot of the smallest 45 eigenvalues for the $H_{21}$ parent Hamiltonian for four particles on a $7 \times 5$ site lattice with a five-site magnetic unit-cell. The 35 machine-zero eigenvalues correspond to the prediction of the EPP-based zero-mode counting.  \label{quasiholespec}}
\end{figure}

As a further test of the validity of the EPP analysis, we examine the ground state degeneracy as an extra flux quantum is added while the particle number is held constant: such an insertion gives the counting of the one-quasihole excitations above the ground state at commensurate filling \cite{Mazaheri_zero_2015}. Consider a lattice pierced by six flux quanta, again at filling fraction
$2/3$, so that at root level the ground states are determined by the
pattern $110110$ and its three translates. Upon inserting one additional
flux quantum, two inequivalent root patterns (up to translation) arise,
namely $1010110$ and $1100110$, each of period seven. Each corresponding
root state therefore forms a sevenfold multiplet under lattice
translations. In addition, the two Landau levels furnish an internal pseudospin
$S=1/2$ degree of freedom that must be taken into account when counting
zero modes. The EPP constrains the allowed pseudospin configurations.
For the pattern $1100110$ (and its translates), the bosonic analog of
the EPP discussed in \cite{chen_jain_2017} requires the occupied
nearest-neighbor orbitals to form pseudospin singlets, yielding no
additional degeneracy beyond translation. By contrast, for the pattern $1010110$ the pseudospins
of the first two occupied orbitals remain unconstrained, producing an
additional fourfold internal degeneracy. Taking both translation and
pseudospin multiplicities into account, the expected number of zero modes
in the presence of the extra flux quantum is therefore
$7\times 1 + 7\times 4 = 35$. In Fig.~\ref{quasiholespec} we present the low-energy spectrum of $H_{21}$ for four particles on a $7\times 5$ site lattice with a five-site unit cell, which again we find to be in agreement with our expected zero-mode counting. Note here that the five-site magnetic unit-cell implies the existence of three degenerate dispersive bands, in distinction from the models with two dispersive bands as displayed in e.g. Fig.~\ref{gsspec}. Had we chosen $m=n=2$ in this case, our lattice with seven flux quanta would necessarily have dimensions $14 \times 2$: we find in general that parent Hamiltonians for lattices with very large or small aspect ratios feature smaller many-body gaps, and that the largest gaps occur for models with aspect ratio $M/N = 1$. 

\subsection{The 22-state -- Ising anyons}

We proceed with a non-Abelian instance of our model, realizing the Jain-22 parton state
\begin{equation}
    \Psi_{22}(z,\bar z) = \chi_2(z,\bar z)^2,
\end{equation}
corresponding to Eq.~\eqref{psi_rr} with $r=2$. This bosonic state occupies $p=3$ Landau levels and occurs at filling factor $\nu=1$.
This state can be recognized as the bosonic partner of the fermionic Jain-221 state \cite{jain_incompressible_1989, jain_theory_1990, wen_non-abelian_1991}, whose parent Hamiltonian enforcing the $\mathcal{M}=3$ clustering condition was studied in detail in \cite{bandyopadhyay_entangled_2018}. A (non-trivial) byproduct of this study is that every zero mode of the Jain-221 parent Hamiltonian is divisible by the holomorphic Jastrow factor $\prod_{i<j}(z_i-z_j)$. It follows that the zero modes for the present, bosonic, 22-state are again in one-to-one correspondence with those of the 221-state, and hence for the genus zero geometries, the zero mode structure can be rigorously inferred from the result of Ref. \cite{bandyopadhyay_entangled_2018}.


Like its fermionic partner, the 22-state is non-Abelian and
hosts Ising-type anyons. As argued above, we therefore expect it to be the unique ground state of $H_{22}$, modulo topological
degeneracy. At the level of root configurations, there are three
inequivalent patterns at bulk filling factor $\nu=1$, namely
$202020\dots$, $020202\dots$, and $1111\dots$.
This implies an expected torus degeneracy of three.

To realize this phase on the lattice, we construct $H_{22}$ with $p=3$ flat bands, again using a magnetic unit-cell with $m=n=2$, yielding three flat bands and a single additional dispersing band. We first examine the single-particle structure of $H_{22}$. The hopping amplitude $t_R = |\braket{R_0|H_{22}|R}|$ is shown in Figs.~\ref{hoppingParameter}, ~\ref{1DhoppingParameter} and ~\ref{hoppingParameter3D} for $18\times18$ and $36\times36$ lattices, while the band structure is given in Fig.~\ref{cutoffBS} for a $40\times40$ lattice. Once again we find that $t_R$ displays exponential decay consistent with the analytic considerations of Sec.~\ref{onebod}, and that the band structure consists of three exactly flat degenerate bands well separated from a single dispersing band. We then compute the many-body spectrum of $H_{22}$ for four particles on a $4\times6$ lattice, in a geometry for which each band contains six single-particle states. The resulting spectrum, shown in Fig.~\ref{gsspec}, exhibits exactly three zero modes, in agreement with the expected torus degeneracy discussed above. Notably, this agreement
already occurs in the presence of only a single dispersing band.



\subsection{The 33-state -- Fibonacci anyons}

As a final example, we consider the parton 33-state,
\begin{equation}
    \Psi_{33}(z,\bar z) = \chi_3(z,\bar z)^2,
\end{equation}
which occupies five Landau levels at filling factor $\nu = 3/2$. 
According to our initial discussion, the 33-state realizes a $SU(2)_3 \times U(1)$ non-Abelian topological phase. The neutral $SU(2)_3$ sector contains Fibonacci-type anyons and is therefore expected to support universal topological quantum computation \cite{freedman_modular_2002,simon_topological_2023}.
This is in part supported by the general root-pattern analysis described earlier, where the 33-state is associated with root patterns $303030\dots$ and (via quasihole insertion) $212121\dots$ at bulk filling factor $\nu=3/2$. On the torus, these patterns produce a ground state degeneracy of four. Moreover, at the level of root {\em patterns}, these findings are the same as for the level $k=3$ bosonic Read-Rezayi state
(at the level of root {\em states}, one would expect to see additional domain-wall pseudo-spin dressings, analogous to the fermionic 222-parton case discussed in \cite{tanhayi_ahari_partons_2023}). Root pattern analysis may then also be used to extract statistical information, as 
discussed in detail for the $k=3$ Read-Rezayi state in \cite{flavin_abelian_2011}.
 
Unlike for the 21- and 22-cases, however, we cannot presently relate the zero modes of the 33-parent Hamiltonian to those of a previously studied fermionic parent Hamiltonian via a simple antisymmetric factor. 
A detailed zero-mode analysis of the 33-parent Hamiltonian therefore remains an interesting problem for the future. 
Nevertheless, the field-theoretic considerations and root-pattern structure strongly indicate a fourfold ground-state degeneracy.

To realize this phase on the lattice, we construct the corresponding parent
Hamiltonian $H_{33}$ with $p=5$ flat bands. In contrast to the $22$ state,
the minimal construction with $q=1$ does not reproduce the expected zero-mode
structure. We therefore consider a larger magnetic unit cell with dimensions
$m=6$, $n=2$, containing 12 sites. In this case the construction produces five
exactly flat zero-energy bands and seven additional dispersing bands. 
Moreover, the hopping
amplitudes $t_R = |\langle R_0|H_{33}|R\rangle|$ again decay rapidly with distance,
as shown in Figs.~\ref{hoppingParameter}, ~\ref{1DhoppingParameter} and ~\ref{hoppingParameter3D} for lattices of size $24\times24$ and $48\times 48$.  The resulting single–
particle band structure is displayed in Fig.~\ref{cutoffBS} for a lattice of size $24\times 24$. In contrast to the previous two cases, the excitation bands of $H_{33}$
are themselves nearly flat. This reflects a general feature of our models:
for large unit-cell size $mn$, or equivalently at low flux
$\phi = 1/mn$, the lattice model approaches the continuum limit in which
the single-particle states $\Psi^{(d)}_{a,b}$ increasingly resemble
continuum Landau-level wavefunctions. In this regime the excited bands
become correspondingly flatter while the single-particle gaps shrink
\cite{Harper_perturbative_2014}. 

\begin{figure}
    \centering
    \includegraphics[width=0.75\linewidth]{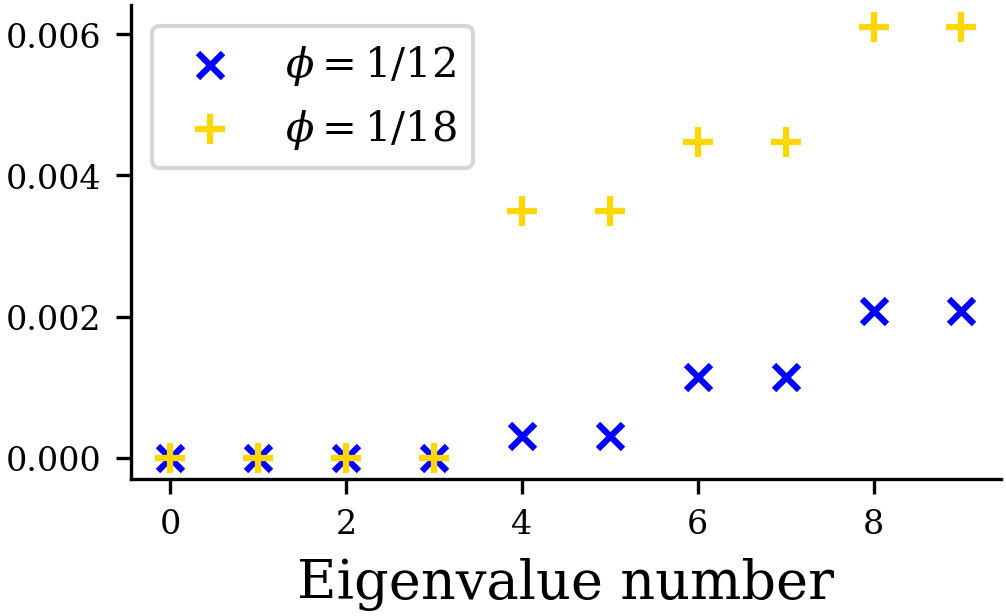}
    \caption{The ten smallest eigenvalues of the 33-state parent Hamiltonian for two values of $\phi = 1/mn$. Results are given for three particles on a lattice of dimension $6 \times 4$ with a 12-site magnetic unit-cell (blue) and $6 \times 6$ with an 18-site unit-cell (gold). In each case the four smallest machine-zero eigenvalues are in agreement with the expected topological ground state degeneracy given by counting root-patterns.\label{33spec}}
\end{figure}

In Fig.~\ref{33spec} we present the low-energy spectrum for three particles
on lattices with two different fluxes: first, a $6\times4$ site lattice
with the 12-site magnetic unit cell introduced above, and second a
$6\times6$ site lattice with an 18-site magnetic unit cell.
In both cases the appearance of precisely four low-lying states,
well separated from higher excitations, confirms the expected torus
degeneracy of the topological phase. We note, however, that the gap is
substantially larger in the lower-flux case.
\subsection{Finite-range parent Hamiltonians}

\begin{figure}[ht]
    \centering
    \includegraphics[width=\linewidth]{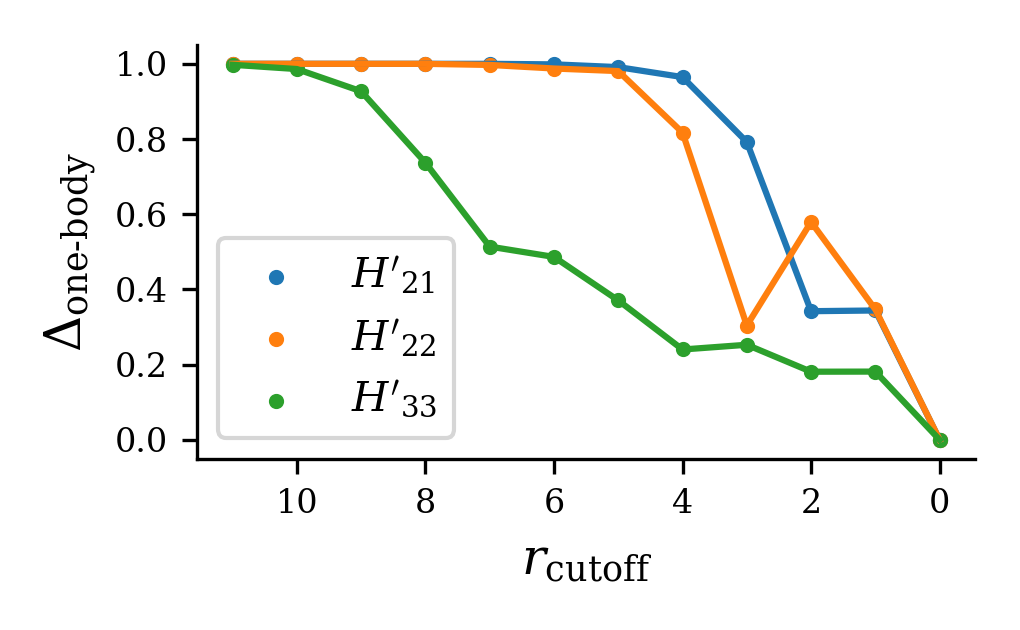}
    \caption{The single-particle spectral gap $\Delta_\text{one-body}$ for the truncated 21-, 22- and 33-lattice state parent Hamiltonians as a function of the cutoff distance $r_\text{cutoff}$  defined in (\ref{rcoeq}). Here we normalize the untruncated parent Hamiltonians $H$ to have unit valued gap. In accordance with Fig~\ref{hoppingParameter}, values are given for lattices of dimensions $14 \times 14$ and $18 \times 18$ with four site magnetic unit-cells for $H_{21}$ and $H_{22}$ and for a $24 \times 24$ site lattice with a 12-site magnetic unit-cell for $H_{33}$.  \label{cutoffGaps}}
\end{figure}

Although our hopping amplitudes decay exponentially in space, they are not strictly finite-ranged, which poses a significant experimental challenge for engineering the model. It is therefore natural to examine how the flat-band structure behaves under truncation. We therefore introduce a truncated  Hamiltonian $H'$ whose matrix elements are given by
\begin{equation}\label{rcoeq}
    \braket{R|H'|R'} = 
    \begin{cases}
        \braket{R|H|R'} & |R-R'| \leq r_\text{cutoff}\\
        0 & |R-R'| > r_\text{cutoff},
    \end{cases}
\end{equation}
for $H$ any of the parent Hamiltonian considered above. In Fig.~\ref{cutoffGaps} we show the resulting single-particle spectral gap between the group of bands evolving from the originally flat bands and the higher dispersing bands as a function of the cutoff radius $r_{\text{cutoff}}$. The gap remains open for moderate truncation radii, indicating that the essential flat-band structure is robust against removal of the longest-range hopping processes.

To further illustrate this stability, Fig.~\ref{cutoffBS} compares the band structures of $H$ and $H'$ with $r_{\text{cutoff}} = $ 3, 4, and 8 for the three models. As expected for a topological flat-band model, truncation necessarily introduces dispersion \cite{chen_impossibility_2014,read_compactly_2017}, which we indeed observe. Nevertheless, for the modest cutoffs considered in Fig.~\ref{cutoffBS}, the dispersion acquired by the group of formerly flat bands remains moderate compared to the gap separating them from the rest of the spectrum.

\section{Conclusion\label{conc}}

In this work, we have constructed a class of exactly solvable lattice Hamiltonians whose single-particle spectrum contains an arbitrary number $p$ of degenerate flat bands that reproduce the analytic structure of the first $p$ Landau levels in the continuum on a torus. 
These bands arise as the gapped ground-state manifold of exponentially localized hopping models, providing an explicit lattice realization of multiple Landau levels with exact analytic control. 
Such constructions offer a useful platform for idealized models of flat-band Chern insulators and for studying correlated phases in systems with higher Chern number.

In the presence of contact interactions, these models stabilize both Abelian and non-Abelian ground states in the universality class of bosonic parton states; here, we gave numerical evidence of this fact in the form of ground-state degeneracies computed for the Jain-$21$, -$22$, and -$33$ states, realized within $p=2$, $3$, and $5$ flat bands, respectively. 
Remarkably, the non-Abelian phases emerge as exact ground states of local two-body interactions. 
In particular, the $33$-state realizes a topological phase expected to host Fibonacci-type anyons, which are capable of universal topological quantum computation. We further emphasize that the ground-state degeneracies and low-energy spectra
obtained from exact diagonalization agree with the predictions of the
entangled Pauli principle and the associated root-pattern analysis, as well
as with the topological field-theory description of the phase.

We further showed that the hopping amplitudes in these models decay exponentially in real space and that truncating the Hamiltonian at moderate ranges $r_{\text{cutoff}}$ preserves the separation between the flat-band manifold and higher bands, although perfect flatness is necessarily lost under such truncation. 
Understanding the stability of the many-body phases under such realistic locality constraints remains an important direction for future work. 
Here, we were not able to perform exact diagonalization at the ``incompressible'' filling factor for lattice dimensions much larger than the relevant range of $r_{\text{cutoff}} \approx 4$--$8$. 
Tracking the evolution of the ground-state manifold and its topological degeneracy as a function of system size and truncation radius would therefore be interesting challenges for future work.

More broadly, the models presented here contain a large parameter space that has not yet been systematically explored. 
Parameters such as the magnetic unit-cell size, the freedom in the choice of coefficients $C_{\vec\beta}$ (and corresponding choice of basis) entering the construction, and the geometry of the lattice within the magnetic unit-cell may provide opportunities to optimize locality, band gaps, and flatness. 
Exploring this parameter space, as well as studying larger system sizes and alternative lattice geometries, may lead to improved realizations of flat-band topological phases.

Finally, we hope that the constructions presented here provide a useful bridge between the analytic structure of continuum quantum Hall physics and lattice realizations of correlated topological matter. 
By realizing multiple Landau levels and non-Abelian phases within exactly solvable two-body lattice models, our work may offer a promising starting point both for theoretical studies of fractional Chern insulators and for future experimental efforts to engineer such phases in  cold-atom systems or other programmable quantum platforms.
\begin{acknowledgments}
AS would like to thank X.-G. Wen, F. Pollmann, and M. Knap for insightful discussions. 
This work was supported in part by the National Science Foundation under Grant No.\ DMR-2029401.
\end{acknowledgments}

\appendix
\section{Degenerate dispersive bands}\label{appA}
We prove that the non-flat bands of equation (\ref{H_one_body}) are degenerate for the special choice of $C_{\vec\beta}\equiv 1$, or equivalently, that in this case $H$ is proportional to a projection operator at each $a$ and $b$. Since our proof will not depend on the values of $a$ and $b$, we will omit them as a subscript. 
The remainder of this appendix makes extensive use of the language of exterior algebras as presented, for example, in Frankel \cite{Frankel_geometry_2011}. Although the pertinent chapters of Frankel are formulated over real vector spaces, we work throughout with complex vector spaces. This extension is standard and requires no essential modification of the definitions, except where explicitly noted below; see, for example, \cite{Greub_multilinear_1967}.

Let $\mathcal H \cong \mathbb C^{mn}$ denote the magnetic unit-cell Hilbert space  as in the main text. Let $\wedge$ denote the exterior (wedge) product over
$\mathbb C$, and let
\[
\Lambda \mathcal H
=
\bigoplus_{i=0}^{\dim \mathcal H} \Lambda^i \mathcal H
\]
denote the exterior algebra of $\mathcal H$.
We furthermore fix a normalized top-degree form $\Omega \in \Lambda^{mn} \mathcal H^*$, satisfying
$\lVert \Omega \rVert = 1$ with respect to the Hermitian norm on $\Lambda^{mn} \mathcal H^*$
induced by the inner product on $\mathcal H$. Since $mn = \dim \mathcal H$, the space
$\Lambda^{mn} \mathcal H^*$ is one-dimensional, and $\Omega$ is therefore unique up to a
phase factor $e^{i\theta}$. As will be clear below, the particular choice of $\Omega$ does
not affect the final results.

To begin, consider the complex anti-linear function $h:\Lambda^{q-1} \mathcal{H} \rightarrow \mathcal{H}$ 
\begin{equation}
    h\of{\ket{\omega}} = \of{i_{ \ket{\Psi^{(0)}} \wedge \dots \wedge  \ket{\Psi^{(p-1)}}  \wedge \ket{\omega} }\Omega}^\dagger
\end{equation}
where $i$ denotes interior multiplication \footnote{This construction may be viewed as a holomorphic generalization of the real Hodge star. Since several inequivalent generalizations of the Hodge star to complex vector spaces appear in the literature, we avoid introducing star notation here.}. Using $h$, we can define a complex linear map $\hat H:\mathcal{H} \rightarrow  \mathcal{H}$ by $\hat H \ket{\psi} = h\circ h^\dagger(\ket{\psi})$ for $\ket{\psi} \in \mathcal{H}$ and $h^\dagger$ the adjoint of $h$. Let $\{\ket{\beta}\}$ be an orthonormal basis of $\mathcal{H}$ and $\{\ket{\vec \beta}\}$ the orthonormal basis of $\Lambda^{q-1}\mathcal{H}$ furnished by taking wedge products of the $\{\ket{\beta}\}$, using some ordering convention. Then $h$ can be written $h = W \hat C$ where $W$ is a complex-linear map and $\hat C$ is the anti-linear unitary (or anti-unitary) operator which implements complex conjugation with respect to the basis $\{\ket{\vec \beta}\}$. It follows that $H$ can be expressed as
\begin{align}\label{Hspectral}
    \hat H  = W \hat C \hat C^\dagger W^\dagger = WW^\dagger = W\of{\sum_{\vec\beta} \ket{\vec\beta}\bra{\vec\beta}} W^\dagger = \sum_{\vec\beta} \ket{h(\vec\beta)}\bra{h(\vec\beta)}
\end{align}
for $\ket{h(\vec\beta)} := W\ket{\vec\beta}$. By construction, this expression is independent of the choice of orthonormal basis $\{\ket{\beta}\}$. 

To make contact with the main text, let $\ket{\beta} \equiv \ket{\Phi^\beta}= \of{\Phi^\beta}_\alpha \ket{e^\alpha}$, where again $\ket{e^\alpha}$ is the normalized ket supported on the site labeled $\alpha$ within the unit cell. Then 
\begin{align}
    \bra{h(\vec\beta)} &= i_{ \ket{\Psi^{(0)}} \wedge \dots \wedge  \ket{\Psi^{(p-1)}}  \wedge \ket{\vec\beta} }\Omega \nonumber \\
        &= \Omega\of{\ket{\Psi^{(0)}},\dots,\ket{\Psi^{(p-1)}},\ket{{\Phi^{\beta^1}}},\dots,\ket{{\Phi^{\beta^{q-1}}}},-} \nonumber \\
        &= \Psi^{(0)}_{\alpha_0}\dots \Psi^{(p-1)}_{\alpha_{p-1}} \of{\Phi_{a,b}^{\beta^1}}_{\alpha_{p}} \dots \of{\Phi_{a,b}^{\beta^{q-1}}}_{\alpha_{mn-2}} \\
        & \ \ \times \Omega\of{\ket{e^{\alpha_0}},\dots,\ket{e^{\alpha_{p-1}}},\ket{e^{\alpha_{p}}},\dots,\ket{e^{\alpha_{mn-2}}},\ket{e^\alpha}} \bra{e_\alpha}. 
\end{align}
For $e^{i\varphi} := \Omega\of{\ket{e^{0}},\dots,\ket{e^{mn-1}}}$, the alternating property of $\Omega$ implies
\begin{align}
\Omega\of{\ket{e^{\alpha_0}},\dots,\ket{e^{\alpha_{mn-2}}},\ket{e^\alpha}} = \epsilon^{\alpha_0\dots\alpha_{mn-2}\alpha} e^{i\varphi}, 
\end{align}
and therefore
\begin{align}
\ket{h(\vec\beta)} =  e^{-i\varphi}\Psi^\perp_\alpha(\vec\beta) \ket{e^{\alpha}} = e^{-i\varphi} \ket{\Psi^\perp(\vec\beta)} 
\end{align}
where Eq. \eqref{psiperpbeta} was used.
Hence up to a phase, $\ket{h(\vec\beta)}$ equals $\ket{\Psi^\perp(\vec\beta)}$ of the main text, and, therefore, the operator $\hat H$ defined here equals
the $k$-space Hamiltonian $H_{a,b}$ as defined in equation (\ref{H_one_body}) with $C_{\vec\beta} \equiv 1$.
Certainly $\hat H\ket{\Psi^{(j)}} = 0$, so what remains is to show that $\hat H$ evaluates to a constant $K$ on vectors orthogonal to the $\ket{\Psi^{(j)}}$. 

We now make use of our freedom to choose a particular basis $\{\ket{\Phi^\beta}\}$. 
By the assumption of linear independence of the $\ket{\Psi^{(j)}}$, the Hilbert subspace $\mathcal{H}^\parallel = \text{span}\of{ \ket{\Psi^{(0)}},\dots,\ket{\Psi^{(p-1)}} }$ has dimension $p$, and admits an orthonormal basis $\{\ket{\tilde \Phi^\beta}\}_{\beta = 0,\dots,p-1}$.
 Clearly $\ket{\Psi} = \ket{\Psi^{(0)}}\wedge \dots \wedge \ket{\Psi^{(p-1)}}$ is a non-vanishing top-degree form on $\mathcal{H}^\parallel$, so without loss of generality we may write $\ket{\Psi} = \sqrt{K} \ket{{\tilde \Phi^{0}}} \wedge \dots \wedge \ket{{\tilde \Phi}^{p-1}}$ with $K := \lVert\ket{\Psi}\rVert^2>0$. Furthermore, extend the basis 
 $\{\ket{\tilde \Phi^\beta}\}_{\beta = 0,\dots p-1}$ to an orthonormal basis
 $\{\ket{\Phi^\beta}\}_{\beta = 0,\dots,mn-1}$ of $\cal H$, with the
  $\{\ket{\Phi^\beta}\}_{\beta = p,\dots,mn-1}$ spanning the orthogonal complement $\cal H^\perp$ of $\mathcal{H}^\parallel$.
Now consider the action of $h$ on the basis vectors of the form $\ket{\Phi^{\vec\beta}} = \ket{{\Phi^{\beta_1}}} \wedge \dots \wedge \ket{\Phi^{\beta_{q-1}}}$. By construction, $h(\ket{\Phi^{\vec\beta}}) = 0$ if ${\beta_j} < p$ for any $1 \leq j \leq q-1$; in contrast, if $\{\beta_1,\dots \beta_{q-1}\} = \{p,\dots,mn-1\} \setminus \{k\}$ for some $p \leq k \leq mn-1$, then we have $h(\ket{\Phi^{\vec\beta}}) = \pm e^{-i\tilde\varphi}  \sqrt{K} \ket{ \Phi^{k}}$,
where $\Omega= e^{i\tilde\varphi} \langle \beta^0|\wedge\dotsc \wedge\langle \beta^{mn-1}|$.
With Eq. \eqref{Hspectral}, this implies 
\begin{equation}
    \hat H = \sum_{\beta=p}^{mn-1} K\ket{\Phi^\beta}\bra{\Phi^\beta}= K P^\perp,
\end{equation}
for $P^\perp$ the projection operator onto $\mathcal{H}^\perp$, which is what we had wanted to show.

\bibliography{main}

\end{document}